\documentclass[12pt]{article}
\usepackage{amsfonts,amssymb,latexsym}
\begin{document}

\centerline{\bf Restoration of Many Electron Wave Functions }

\centerline{\bf from One-Electron Density }

\centerline {A. I. Panin$^1$ and A. N. Petrov$^2$}

\centerline{ \sl $^1$Chemistry Department, St.-Petersburg State University,}
\centerline {\sl University prospect 26, St.-Petersburg 198504, Russia }
\centerline { \sl e-mail: andrej@AP2707.spb.edu }

\centerline{ \sl $^2$Petersburg Nuclear Physics Institute, Gatchina,}
\centerline {\sl St.-Petersburg District 188300, Russia }
\sloppy

\bigbreak
{\bf ABSTRACT: }{\small

General theorem describing a relation between diagonal of one-electron
density matrix and a certain class of many-electron ensembles of determinant
states is proved. As a corollary to this theorem a constructive proof of
sufficiency of Coleman's representability conditions is obtained.
It is shown that there exist rigorous schemes for construction of
energy of many-electron system as functionals of one-electron density.
}

\bigbreak

\bigbreak
\hrule
\bigbreak
{\bf Introduction}

In density functional theory (DFT) approaches it is accepted that
the electronic energy of many electron systems (at least for the ground state)
can be presented as a functional of the first order density, or, in other
words, of the diagonal of the first order density matrix. In this connection the
following question seems to be pertinent: Are there rigorous schemes,
that do not involve approximations and hypothesis of any kind,
for construction of electronic energy as a functional of the first
order density ? In present paper we make an attempt to answer this question.

The second question closely related to the first one may be formulated as:
Is it possible to
consider electronic energy of many-electron system for some fixed basis as a
function of occupation numbers ? Recently orbital occupancy (OO) approach
developing this idea for DFT type functionals was formulated [1-4].
In present
paper we show how energy expressions involving diagonal elements of the first
order density matrix may be rigorously constructed. Such energy expressions
may be used only in non-gradient optimization schemes since these expressions
are not differentiable in classic sense.

\bigbreak
\bigbreak
\hrule
\bigbreak

{\bf Basic Definitions}
\bigbreak

For a fixed basis set of $n$ orthonormal spin-orbitals
the corresponding finite-dimensional
Fock space ${\cal F}_N$ is  spanned by determinants
$|L\rangle $ where $L$ runs over all subsets of the spin-orbital index set $N$.
Its $p$-electron sector ${\cal F}_{N,p}$ is spanned by determinants
$|L\rangle $ with $|L|=p$.
Basis determinants will be  labelled by {\sl subsets}   and  all sign conventions
connected with their  representation as  the Grassman
product of {\sl ordered} spin-orbitals will be included in the definition of the
creation-annihilation operators.

The set of $q$-electron density  operators is defined as
$${\cal E}_{N,q}=\{t_q\in {\cal F}_{N,q} \otimes {\cal F}_{N,q}^*:
t_q^{\dagger}=t_q \ \& \ t_q \ge 0 \ \&\  Tr (t_q)=1 \}\eqno (1)$$
The diagonal mapping over ${\cal F}_N \otimes {\cal F}_N^*$ is
$$d(t)
=\sum\limits_{L\subset N} \langle L|t|L \rangle e_L,
\eqno(2)$$
where $t\in {\cal F}_N \otimes {\cal F}_N^*$
and $e_L=|L \rangle \langle L|$.

The contraction operator over ${\cal F}_N \otimes {\cal F}_N^*$ is defined
in terms of the standard fermion creation-annihilation operators as
$$c=\sum\limits_{i=1}^na_i\otimes a_i^{\dagger}.\eqno(3)$$

{\bf Definition 1.} q-electron density operator $t_q\in {\cal E}_{N,q}$ is called
weakly $p$-representable if there exists $p$-electron density operator
$t_p\in {\cal E}_{N,p}$ such that
$$ \frac {q!}{p!}c^{p-q}(d(t_p))=d(t_q).\eqno(4)$$
This definition is correct because the contraction operator possesses the
property
$$c(d({\cal E}_{N,p}))\subset d({\cal E}_{N,q}).$$

The set $d({\cal E}_{N,p})$ is called the standard (unit) simplex
of the operator space $d({\cal F}_{N,p} \otimes {\cal F}_{N,p}^*)$
and its characterization is given by
$${\cal T}_{N,p}=d({\cal E}_{N,p}) = \{
\sum\limits_{L\subset N}^{(p)}{\lambda }_Le_L:{\lambda }_L\ge 0 \ \&
\ \sum\limits_{L\subset N}^{(p)}{\lambda}_L=1\}.
\eqno(5)$$

The combinatorial structure of ${\cal T}_{N,p}$ is very simple: Any part
of the set of all $p$-element subsets of the index set N
determines a face of ${\cal T}_{n,p}$ and its complementary part generates
the opposite face. In particular, there are $n\choose p$ hyperfaces opposite to
the corresponding vertices.

{\bf Definition 2.}
$$W_{N,p,q}=\frac {q!}{p!} c^{p-q}d({\cal E}_{N,p}).\eqno(6)$$

Weak representability problem may be formulated as the problem of
description of the polyhedron $W_{N,p,q}$ with arbitrary
admissible $n,p$, and $q$. Since, by definition, density operators
are Hermitean, this polyhedron may be conveniently embedded into
the {\it real} Euclidean space ${\mathbb R}_{N,q}$ of the
dimension ${n \choose q}$ with its canonical basis vectors $e_K$
labelled by $q$-element subsets of $N$. With such an embedding the
tensor products of the fermion creation-annihilation operators
involved in the expression (3)  should be replaced by the
commuting (Bose) annihilation operators
$$b_je_J= \cases {e_{J\backslash \{j\}},&if $j\in J$\cr
           0,&if $j\notin J $ \cr},\eqno(7)$$

acting on the vector space
$${\mathbb R}_N=\bigoplus\limits_{q=0}^n {\mathbb R}_{N,q}.\eqno(8)$$

$W_{N,p,q}$ is a polyhedron situated in the real affine hyperplane
$${\cal H}_{N,q}=\{\lambda \in {\mathbb R}_{N,q}:
\sum\limits_{J\subset N}^{(q)}{\lambda }_J=1\}.\eqno(9)$$

Let us put
$$w_{p\downarrow q}(L)=\frac {q!}{p!}c^{p-q}e_L=\frac{1}{{p\choose q}}
\sum\limits_{K\subset L}^{(q)}e_K.\eqno(10)$$

Directly from definition it follows that the polyhedron $W_{N,p,q}$
is the convex hull of ${n\choose p}$ vectors $w_{p\downarrow q}(L)$:

$$W_{N,p,q}=Conv(\{w_{p\downarrow q}(L)\}_{L\subset N}). \eqno(11)$$
To the best of our knowledge, in contrast to the parametric description
given by Eq.(11), the analytic description (that is the description in
terms of the hyperfaces) of this polyhedron is obtained only for the
case $q=1$ and is given by the following assertion that is just a consequence
of the general theorem by Coleman [5-10]:

{\bf Theorem 1.} Polyhedron $W_{N,p,1}$ is the set of solutions of the system
$$\cases {
0\le {\lambda}_k\le \frac {1}{p},\ k\in N \cr
\sum\limits_{j\in N}{\lambda}_j=1\cr
}\eqno(12)$$

This polyhedron has $2n$ hyperfaces with normals
$$v_k^0=pe_k,\eqno(13a)$$
and
$$v_k^1=-pe_k+\sum\limits_{j\in N}e_j,\eqno(13b)$$
where $k\in N$, and $e_k$ are canonical basis vectors of the
Euclidean space ${\mathbb R}^n$.

\bigbreak
\hrule
\bigbreak
{\bf Restoration of p-Electron Wave Functions from One-Electron}

{\bf  Density Matrix Diagonal}
\bigbreak

With arbitrary vector ${\lambda}^{(0)} \in  W_{N,p,1}$ it is
convenient to associate two index sets:
$$Ind({\lambda}^{(0)})=\{i\in N: {\lambda}^{(0)}_i>0\}\eqno(14a)$$
$$Ind_{\frac{1}{p}}({\lambda}^{(0)})=\{i\in N: {\lambda}^{(0)}_i=\frac{1}{p}\}\eqno(14b)$$

Let us present vector ${\lambda}^{(0)} \in  W_{N,p,1}$ as the convex combination
$${\lambda}^{(0)} =p{\mu}^{L_0}w_{p\downarrow 1}(L_0)+(1-p{\mu}^{L_0}){\lambda}^{(1)}
\eqno(15)$$
where (see Eq.(10))
$$w_{p\downarrow 1}(L_0)=\frac{1}{p}
\sum\limits_{i\in L_0}e_i,\eqno(16)$$

$${\lambda}^{(1)}=\sum\limits_{i \in L_0}\frac{{\lambda}^{(0)}_i-{\mu}^{L_0}}
{1-p{\mu}^{L_0}}e_i+\sum\limits_{i \in N\backslash L_0}
\frac{{\lambda}^{(0)}_i}{1-p{\mu}^{L_0}}e_i,\eqno(17)$$
and require the residual
vector  ${\lambda}^{(1)}$ to be representable. This requirement imposes
the following restrictions on the admissible values of parameter ${\mu}^{L_0}$:
$$\cases {
0\le \frac{{\lambda}^{(0)}_i-{\mu}^{L_0}}{1-p{\mu}^{L_0}}\le \frac{1}{p},\  i\in L_0\cr
0\le \frac{{\lambda}^{(0)}_i}{1-p{\mu}^{L_0}}\le \frac{1}{p},\ i\in N\backslash L_0\cr
}\eqno(18)$$
The frontier solution  of system (18) is
$${\mu}^{L_0}=\min\{\min_{i\in L_0}\{{\lambda}^{(0)}_i\},
\min_{i\in N \backslash
L_0}\{\frac{1}{p}-{\lambda}^{(0)}_i\}\}.\eqno(19)$$ If
${\mu}^{L_0}\ne 0$ then we arrive at non-trivial representation of
diagonal ${\lambda}^{(0)}$  as a convex combination of vertex
$w_{p\downarrow 1}(L_0)$ and a certain representable residual
vector ${\lambda}^{(1)}$. From Eq.(19) it is easy to see that the
additional condition ${\mu}^{L_0}\ne 0$ holds true if and only if
subset $L_0$ satisfies the restriction
$$ Ind_{\frac{1}{p}}(\lambda^{(0)})\subset L_0 \subset Ind(\lambda^{(0)})
\eqno(20)$$

Iterating of Eq.(15) leads to the following expression
$${\lambda}^{(0)}=\sum\limits_{i=0}^{k-1}\left [\prod\limits_{j=0}^{i-1}
(1-p{\mu}^{L_j})\right ]p{\mu}^{L_i}w_{p\downarrow 1}(L_i)+
\left[\prod\limits_{i=0}^{k-1}(1-p{\mu}^{L_i})\right ]{\lambda}^{(k)}\eqno(21)$$
where
$${\mu}^{L_i}=\min\{\min_{l\in L_i}\{{\lambda}^{(i)}_l\},
\min_{l\in N \backslash L_i}\{\frac{1}{p}-{\lambda}^{(i)}_l\}\}\eqno(22)$$
and
$$ Ind_{\frac{1}{p}}(\lambda^{(i)})\subset L_i \subset Ind(\lambda^{(i)})
\eqno(23)$$
for $i=0,1,\ldots,k-1$.

{\bf Definition 3.} Sequence $(L_0,L_1,\ldots,L_i,\ldots)$ of
$p$-element subsets of $N$ is called $\lambda$-admissible if for
each $i=0,1,\ldots$ subset $L_i$ satisfies the condition (23).

{\bf Theorem 2.} For any vector ${\lambda}^{(0)} \in  W_{N,p,1}$
the residual vector in iteration formula (21) vanishes after
a finite number of steps.

{\bf Proof.} First let us note that the number of nonzero components of
representable residual vector
${\lambda}^{(k)}$
can not be less than $p$. If this number is equal to $p$ then
${\lambda}^{(k)}$  just coincides with the
vertex $w_{p\downarrow 1}(L_k)$ where $L_k=Ind({\lambda}^{(k)})$,
and the residual vector ${\lambda}^{(k+1)}$ vanishes. Let us suppose that
the number of nonzero components of ${\lambda}^{(k)}$ is greater than $p$.
From Eqs.(15), (17),
and (19) it readily follows that there exists index $i_*\in Ind({\lambda}^{(k)})$
such that  ${\lambda}^{(k+1)}_{i_*}$ is necessarily equal either to zero or to
$\frac{1}{p}$. To complete the proof it is sufficient to show that if
${\lambda}^{(k)}_{i}=\frac{1}{p}$ then ${\lambda}^{(k+1)}_{i}=\frac{1}{p}$.
Condition (23) implies that all the indices $i\in N$ such that
${\lambda}^{(k)}_i=\frac{1}{p}$ should belong to $L_k$ because in the opposite
case the parameter ${\mu}^{L_k}$ would be equal to zero.
If ${\mu}^{L_k}={\lambda}^{(k)}_{i_*} > 0$ and
${\lambda}^{(k)}_{i}=\frac{1}{p}$
then
${\lambda}^{(k+1)}_{i}=\frac{\frac{1}{p}-{\lambda}^{(k)}_{i_*}}{1-p{\lambda}^{(k)}_{i_*}}
=\frac{1}{p}$. If, on the other hand,
${\mu}^{L_k}=\frac{1}{p}-{\lambda}^{(k)}_{i_*} > 0$ and
${\lambda}^{(k)}_{i}=\frac{1}{p}$
then
$1-p{\mu}^{L_k}=p{\lambda}^{(k)}_{i_*}$ and
${\lambda}^{(k+1)}_{i}=\frac{\frac{1}{p}-{\mu}^{L_k}}{p{\lambda}^{(k)}_{i_*}}
=\frac{1}{p}$
$\Box $

{\bf Corollary 1.} The set of solutions of the Coleman's system (12) is the
convex hull of ${n\choose p}$ vertices $w_{p\downarrow 1}(L)$.

{\bf Corollary 2.} The number of vertices in expansion of a given
density diagonal  obtained on the base of the recurrence formula (21)
is not greater than the number of its components
different from zero.

{\bf Corollary 3.} $\lambda$-admissible sequence $(L_0,L_1,\ldots,L_{k_{\lambda}})$
generated recurrently on the base of the iteration formula (21) includes pairwise
distinct $p$-element subsets and
$$\bar{\lambda}(L_0,L_1,\ldots,L_{k_{\lambda}}) =
\sum\limits_{i=0}^{k_{\lambda}}\left [\prod\limits_{j=0}^{i-1}
(1-p{\mu}^{L_j})\right ]p{\mu}^{L_i}e_{L_i}\eqno(24)$$
is a diagonal of $p$-electron density matrix such that
$$\frac{1}{p!}c^{p-1}\bar{\lambda}(L_0,L_1,\ldots,L_{k_{\lambda}})= \lambda.\eqno(25)$$
It is to be noted that Theorem 2 is just a specification of the
fundamental theorem by Carath\'eodory \cite {Carat}:

{\bf Theorem 3.} Let $X\subset {\mathbb R}^n$. Then any vector
$x\in Conv(X)$ may be presented as a convex combination of no more
than $n+1$ vectors from $X$.
Modern proof of this result may be found in \cite {Leicht}.

From Corollary 3 if follows that any mapping $\lambda \to s_{\lambda}$
where $s_{\lambda}$ is a $\lambda$-admissible sequence compatible with
the iteration formula (21) determines some global section (right inverse)
${\pi}_{1\uparrow p}$ of the contraction operator $\frac{1}{p!}c^{p-1}$
that is the mapping from $W_{N,p,1}$ to
${\cal T}_{N,p}$ such that
$$\frac{1}{p!}c^{p-1}{\pi}_{1\uparrow p}(\lambda)=\lambda \eqno(26)$$
for any $\lambda \in W_{N,p,1}$. As it is seen from Eq.(22), sections
constructed on the base of the recurrence relation (21) are not linear and even
not differentiable in classic sense.

The most ambitious task arising in the frameworks of the approach outlined
is to try to develop efficient methods for direct optimization of energy as
a function of diagonal of the first order density matrix. General scheme
embracing the whole class of such methods may be described as follows.

1. Some section(s) of the contraction operator should be chosen.

2. Using available section, it is possible of associate with some trial
diagonal $\lambda$ ensemble of $p$-electron determinant states and to
determine squares of the CI coefficients:
$$|C_{L_i}|^2=\left [ \prod\limits_{j=0}^{i-1}(1-p{\mu}^{L_j})\right ]p{\mu}^{L_i}\eqno(27)$$
(see Eq.(24)).

3. Construct average energy
$$E_{\lambda}({\phi}_0,{\phi}_1,\ldots,{\phi}_{k_{\lambda}})=
\sum\limits_{i,j=0}^{k_{\lambda}}cos({\phi}_i-{\phi}_j)|C_{L_i}||C_{L_j}|
<L_i|H|L_j>\eqno(28)$$
as a function of phases ${\phi}_i$.

4. Minimize the function
$$E_{{\pi}_{1\uparrow p}}:\lambda \to   \min_{\phi}E_{\lambda}({\phi})\eqno(29)$$
to determine optimal diagonal and its  expansion
via vertices $w_{p\downarrow 1}(L)$.

There are no serious problems in implementation of steps 2-4 of
this scheme and the only complicated step is reasonable selection
of mapping(s) ${\pi}_{1\uparrow p}$ (note that in general several
different sections may be employed in the course of the energy
optimization). It is rather difficult to estimate {\it a priori}
the quality of some chosen concrete section ${\pi}_{1\uparrow p}$.
There are two readily coming to mind general algorithms to
construct such sections. Both of them involve full sorting of
$p$-electron subsets of the spin-orbital index set.

1. Maximization of parameter (22) on each iteration: On the k-th
step current $L_k$ may be determined from the condition
$${\mu}^{L_k}=\max_{L}\left \{\min\{\min_{l\in L}\{{\lambda}^{(k)}_l\},
\min_{l\in N \backslash
L}\{\frac{1}{p}-{\lambda}^{(k)}_l\}\}\right \}.\eqno(30)$$ In this
case it is not necessary to take into account Eq.(23) explicitly.

This section is probably optimal from formal mathematical
viewpoint but has no physical idea behind it. Computer experiments
show that in restoration process of such type high order
excitations from the HF state contribute mostly. Even if exact FCI
occupancies for the ground state are chosen, the restoration
produces ensemble of determinant states that involves HF
determinant and excited determinants that practically do not
interact with the HF one.

2. Energy minimization:
On the k-th iteration among subsets satisfying the condition (23) it is chosen the
subset $L_k$ such that the lowest eigenvalue of $p$-electron  Hamiltonian
in the basis $\{|L_0>,|L_1>,\ldots,|L_k>\}$ is minimal.

This is undoubtedly the best possible section of the contraction operator.
Unfortunately, the use of this section for the energy
minimization is of no sense because it is equivalent to a certain CI
scheme that can be described as follows.

1. First it is necessary to fix the maximal number $mdet$ of determinants in
wave function expansion and put $k=0$;

2. Put $k=k+1$;

3. Sort all
determinants different from the already chosen  and select the one that
corresponds to the lowest eigenvalue of the Hamiltonian in the basis of $k$
determinants $\{|L_1>,|L_2>,\ldots,|L_k>\}$. If $k<mdet$, return to step 2.

Finally in a certain sense optimal basis involving not greater than
$mdet$ determinants will be obtained. This scheme is based on the well-known
bracketing theorem of matrix algebra (see, e.g.,\cite {W}) and is used in quantum
chemistry for years in different modifications to select initial determinant
space for multi-reference CI calculations \cite {Buenker-1,Buenker-2}.
In our opinion this scheme
is interesting in its own right as a self-sufficient one when a relatively small
number of leading determinants should be constructed from active orbitals
with close orbital energies (the case that occurs extensively in transition metal
complexes) because

(1) CI spaces of huge dimensions can be efficiently handled and
disk memory usage is minimal;

(2) Calculations can be easily restarted;

(3) Algorithms are trivially parallelized  and
if, say, PC clusters are used,  data transfer via local net is minimal;

(4) It is easy to handle both single excited state and a group of
successive states.

For the restoration purpose the above scheme can be considered as
a certain benchmark one because it gives the best possible
occupancies and energy that can be obtained on the basis of the
restoration routine described by Theorem 2.

\bigbreak
\hrule
\bigbreak
{\bf Conclusion }

General theorem establishing a connection between diagonal of the
first order density matrix and a certain set of many-electron wave
functions is proved. It is shown that rigorous energy expression
involving only one-electron density  becomes well-defined as soon
as a certain right inverse of the contraction operator is chosen.
For a fixed representable diagonal of the first order density
matrix there exist quite a number of ways to restore $p$-electron
determinant ensembles that are contracted to the diagonal under
consideration. Each such way is in fact a path of a rather
complicated graph with its vertices labelled by admissible (in
sense of definition 3) $p$-element spin-orbital index sets. The
main problem arising in implementation of optimization schemes
based on such energy expressions is the lack of general simple
algorithms for selection of admissible paths for restoration of
wave functions from one-electron densities. Such algorithms,
besides requirement being simple, should generate paths close in a
certain sense to ones obtained by the benchmark calculations based
on the bracketing theorem. Search for such algorithms is in
progress now. Note in conclusion that the recurrence formula (21)
can be easily generalized to treat densities of higher order and
the only obstacle here is the lack of the complete set of
inequalities for analytic description of the polyhedron
$W_{N,p,q}$ in the case $q>1$. \bigbreak \bigbreak \hrule
\bigbreak

{\bf ACKNOWLEDGMENTS}
\bigbreak
One of us (ANP) gratefully acknowledges the Ministry of Education of the
Russian Federation (Grant PD 02-1.3-236) and the St. Petersburg Committee on
Science and Higher Education (Grant PD 03-1.3-60)
for financial support of the present work.

\bigbreak
\hrule
\bigbreak

 \end{document}